\documentclass[apj]{emulateapj}

\slugcomment{\it Accepted for publication in ApJ}
\shortauthors{Comerford, Moustakas, \& Natarajan}
\shorttitle{Formation Histories of Strong Lensing Clusters}

\begin{document}

\title{Observed Scaling Relations for Strong Lensing Clusters: Consequences for Cosmology and Cluster Assembly}

\author{Julia M. Comerford\altaffilmark{1},  Leonidas
  A. Moustakas\altaffilmark{2}, and
  Priyamvada Natarajan\altaffilmark{3,4}}

\affil{$^1$Astronomy Department, 601 Campbell Hall, University of
California, Berkeley, CA 94720}
\affil{$^2$Jet Propulsion Laboratory, California Institute of
  Technology, MS 169-327, 4800 Oak Grove Drive, Pasadena, CA 91109}
\affil{$^{3}$Department of Astronomy, Yale University, P. O. Box 208101, New Haven, CT 06511}
\affil{$^{4}$Radcliffe Institute for Advanced Study, Byerly Hall, 8 Garden Street, Cambridge, MA 02138}

\begin{abstract}
Scaling relations of observed galaxy cluster properties are useful tools for constraining cosmological parameters as well as cluster formation histories.  One of the key cosmological parameters, $\sigma_8$, is constrained using observed clusters of galaxies, although current
estimates of $\sigma_8$ from the scaling relations of dynamically relaxed galaxy clusters are limited by the large scatter in the observed cluster mass-temperature ($M-T$) relation.  With a sample of eight strong lensing clusters at $0.3 < z <0.8$, we find that the observed cluster concentration-mass relation can be used to reduce the $M-T$ scatter by a factor of 6.  Typically only relaxed clusters are used to estimate $\sigma_8$, but combining the cluster concentration-mass relation with the $M-T$ relation enables the inclusion of unrelaxed clusters as well.  Thus, the resultant gains in the accuracy of $\sigma_8$ measurements from clusters are twofold: the errors on $\sigma_8$ are reduced and the cluster sample size is increased. Therefore, the statistics on $\sigma_8$ determination from clusters are greatly improved by the inclusion of unrelaxed clusters. 
Exploring cluster scaling relations further, we find that the correlation between brightest cluster galaxy (BCG) luminosity and cluster mass offers insight into the assembly histories of clusters.  We find preliminary evidence for a steeper BCG luminosity - cluster mass relation for strong lensing clusters than the general cluster population, hinting that strong lensing clusters may have had more active merging histories.
\end{abstract}

\keywords{ cosmological parameters -- clusters: individual (3C~220, A~370, Cl~0024, Cl~0939, Cl~2244, MS~0451, MS~1137, MS~2137) -- dark matter -- galaxies: evolution -- galaxies: formation -- gravitational lensing }

\section{Introduction}
\label{intro}

As the most massive bound systems known, galaxy clusters provide an important link in understanding the composition and growth of structure in the Universe.  Clusters follow a variety of observational scalings of mass with temperature, luminosity, or cluster counts, and these scalings are sensitive to cosmological parameters including the matter density parameter $\Omega_{\rm m}$, the cosmological constant density parameter $\Omega_\Lambda$, the dark energy equation-of-state parameter $w$, and the normalization of the matter power spectrum $\sigma_8$ (e.g., \citealt{HA01.2,BA02.2,LE02.1,SC03.2,AL04.2,VI09.1}).  Such constraints from galaxy clusters complement the constraints on cosmological parameters from Type Ia supernovae and cosmic microwave background observations.

However, useful galaxy cluster constraints on cosmological parameters
depend primarily on accurate determinations of cluster masses.  Observationally,
cluster masses are typically measured in one of three ways.

A long-established method for determining cluster masses employs the
virial theorem and the measurement of 
velocities of the galaxies that constitute the cluster.  Based on the
three assumptions that the cluster is in virial equilibrium, the galaxy distribution efficiently traces
the cluster mass distribution, and the velocity dispersions
$\sigma$ of the galaxies are isotropic, the cluster mass contained within a radius
$r$ is estimated $M \sim \sigma^2 r/G$.  However, these mass estimates
may be biased as a result of galaxy velocity anisotropies or if
the galaxy distribution does not follow the total mass
distribution (e.g., \citealt{BA82.1}).    

A second method uses cluster X-ray emission as a tracer of cluster masses.
The hot intracluster gas, which is the dominant baryonic component of a cluster and is typically twice the mass of the total mass of the galaxies
in a cluster, emits X-rays via bremsstrahlung radiation and atomic
line emission.  With the temperature $T$ and radial density 
$\rho(r)$ profiles determined
from X-ray spectra and surface brightness distributions, the cluster
mass is given by $M \sim r^2/\rho(r) \; d(-\rho T)/ dr$.  This method assumes
that the intracluster gas is spherically distributed and is in hydrostatic
equilibrium \citep{EV96.1}.  However, these assumptions may be
incorrect. If the gas distribution is not spherical, X-ray mass estimates will be biased by projection effects.  Many galaxy clusters are also not in hydrostatic equilibrium, in particular dynamically unrelaxed clusters that are
undergoing mergers.  There is evidence that the bias of hydrostatic equilibrium mass is linked to the dynamical state of the galaxy cluster (e.g., \citealt{AN09.1,ZH09.1}). In
addition, the hot gas of galaxy clusters with buoyant bubbles near their cores might indicate a departure from hydrostatic equilibrium (e.g.,
\citealt{CH01.1}). 

The most direct estimates of cluster masses employ gravitational lensing
distortions of background galaxies.  This technique is free of 
assumptions about the dynamical state of the cluster, which enables it in
principle to yield more consistent mass estimates, though it is also sensitive to projection effects.  More accurate
cluster mass
estimates can in turn provide tighter constraints on cosmological parameters, and therefore it is of key importance to reduce the errors in cluster mass estimates. 

For example, the primary source of error in cluster-based determinations of
$\sigma_8$ is the error in the mass-temperature relation for relaxed
clusters (e.g., \citealt{PI03.1,HE04.2,VO05.1}).  Recent studies show that an X-ray independent mass approach such as gravitational lensing provides a unique tool to calibrate the mass - temperature relation (e.g., \citealt{SM05.1, MA08.1, ZH08.1}).
Here, we use strong gravitational lensing mass measurements of a sample of eight strong
lensing clusters at $0.3 < z 
< 0.8$ to accurately measure the galaxy cluster mass-temperature
relation.  We also include the effects of cluster concentrations in an
effort to further reduce the scatter in the cluster mass-temperature
relation, which would ultimately enable tighter constraints on 
$\sigma_8$.

In addition to the correlations that exist between cluster properties,
some observational properties of brightest cluster galaxies (BCGs) also
scale with 
properties of the host clusters.  Whereas scalings between cluster
properties are
sensitive to
cosmological parameters, scalings between BCGs and their host clusters
provide constraints on BCG
formation and the evolution of clusters.  

BCGs are a unique population: they are the most massive and luminous galaxies in the Universe.  They are typically located near the centers of clusters, which suggests that a BCG's formation history is intricately linked to the formation of the cluster itself. However, the formation of BCGs is still poorly understood.  

BCGs may form after their host clusters assemble in one of two ways.  First, a
BCG may be the first galaxy to be dragged in by dynamical friction to the center of the dark
matter halo destined to become a cluster, where it then grows through galactic cannibalism by
merging with subsequent galaxies that fall to the center (e.g.,
\citealt{OS75.1,HA78.1}).  However, this scenario typically requires more than a
Hubble time to form a BCG because much of the mass of the infalling galaxy is tidally
stripped, which reduces the dynamical friction effect and slows the
infall \citep{ME85.1}.  

BCG formation may also occur after cluster formation if the host cluster's
central cooling flow forms stars at the cluster center and those stars
build the BCG \citep{CO77.1}. There are several instances of ongoing or recent star formation in BCGs that occupy cooling-flow clusters (e.g., \citealt{CA98.1,CR99.1, HI05.1,MC06.1}), but it is unclear whether the star formation is fueled by the cooling flows or by cold gas brought in through recent galaxy mergers \citep{BI08.2}.

In another scenario, BCGs might form in concert with their host clusters.
A BCG may begin with several galaxies merging together in a group to
form a large galaxy, and then when groups merge as hierarchical
structure formation continues, this large galaxy eventually becomes a
BCG in a massive cluster (e.g., \citealt{ME85.1, DU98.1, BO06.1}).  

Here, we examine the correlation between BCG luminosity and cluster
mass in eight strong
lensing clusters at $0.3 < z 
<0.8$.  This will enable constraints not only on BCG and cluster formation in general,
but also on how the BCGs in strong lensing clusters may have formed
and evolved differently than BCGs in the general cluster population.

The rest of this paper is organized as follows.
In Section~\ref{sample} we describe the selection of our cluster sample, and Section~\ref{properties} gives the masses, dynamical states, and X-ray temperatures for these clusters.  In Section~\ref{mtrelation} we find the $M-T$ relation for the relaxed clusters in our sample and show how the inclusion of cluster concentrations both significantly reduces the scatter in the $M-T$ relation and lifts the restriction on cluster dynamical state.  In Section~\ref{bcgprop} we identify the BCGs in our sample and measure their luminosities. We use these luminosities in Section~\ref{lm} to measure the correlation between BCG luminosity and cluster mass, and we find preliminary evidence that strong lensing clusters may have more active merging histories than the general cluster population.  Section~\ref{conclude} presents our conclusions.
Throughout this paper, we adopt a spatially flat cosmological model
dominated by cold dark matter and a cosmological constant ($\Omega_{\rm m}=0.3$, 
$\Omega_\Lambda=0.7$, $h=0.7$). 

\section{Sample Selection}
\label{sample}

We base our sample on 10 well-known strong lensing clusters
analyzed in \cite{CO06.1}.  All 10 clusters have {\it Hubble Space
  Telescope} ({\it HST}) imaging, which make possible the mass determinations
and photometry measurements central to this paper.  However,
there are no published arc redshifts for two of the clusters, Cl~0016$+$1609
and Cl~0054$-$27, which limits the strong lensing determination of their
cluster masses to the unknown factor $D_{\rm s}/D_{\rm ls}$, the ratio of the
angular diameter distances to the source and between the lens and
source.  Consequently we remove these two clusters, and our sample consists of the
remaining eight clusters at $0.3 < z < 0.8$: ClG~2244$-$02, Abell~370, 3C~220.1,
MS~2137.3$-$2353, MS~0451.6$-$0305, MS~1137.5$+$6625, Cl~0939$+$4713,
and ZwCl~0024$+$1652.  

\section{Cluster Properties}
\label{properties}

Strong correlations are found between cluster observables, and the resultant scaling relations clearly encapsulate key information about cosmological parameters and the assembly history of clusters.  Cluster masses are a component of many cluster scaling relations, and we measure strong lensing masses for our sample of clusters and
compare these to mass estimates from the distributions
of cluster X-ray gas.  Based on
these comparisons and other observable properties of the cluster, we
determine the dynamical state of each cluster as relaxed or
unrelaxed.  We also present cluster X-ray temperatures, which are another component of cluster scaling relations.

\subsection{Cluster Strong Lensing Mass Determination}
\label{mass}

We model each cluster mass distribution with an elliptical Navarro-Frenk-White
(NFW; \citealt{NA96.1, NA97.1}) dark
matter halo centered on the BCG, using the best-fit NFW parameters found by \cite{CO06.1}.   
Strong lensing
arcs with measured redshifts observed in a cluster constrain its mass distribution, and
\cite{CO06.1} use the arcs to characterize best-fit NFW ellipsoids
to each cluster.  With the NFW dark matter
halos completely defined in this way, we can determine any
cluster radius
$r_{\Delta}$ as the radius at which the density of the halo is $\Delta$ times
the critical density at the cluster redshift.

Lack of information about the clusters' three-dimensional shapes
prevents us from calculating their elliptical masses, but
instead we determine the equivalent mass of a spherical NFW halo.
With the \cite{CO06.1} best-fit scale convergence $\kappa_\mathrm{s}$
and scale radius $r_\mathrm{s}$, we estimate the cluster mass within radius $r_\Delta$ as

\begin{equation}
 M_\Delta=4\pi\Sigma_\mathrm{crit}\kappa_\mathrm{s}r_\mathrm{s}^2\,
  \left[\ln(1+x)-\frac{x}{1+x}\right]\;, 
\end{equation} 
where $x \equiv r_\Delta/r_\mathrm{s}$ and $\Sigma_\mathrm{crit}$ is
the critical surface mass density, defined as 
\begin{equation}
\Sigma_\mathrm{crit} \equiv \frac{c^2} {4 \pi G} \frac{D_{\rm s}} {D_{\rm l}
  D_{\rm ls}} \; ,
\end{equation}
which depends on the angular diameter distances $D_{\rm l,s,ls}$ from the
observer to the lens, to the source, and from the lens to the source,
respectively. 

We estimate the errors in mass by propagating the errors in the
best-fit NFW parameters.  As detailed in \cite{CO06.1} these errors
are quite small but are realistic, because the reproduced lensed
image is sensitive to slight variations in a parameter's value.
However, we note that these errors are relevant only to the choice of
lens model and data and do not represent a global systematic uncertainty.

We use the method described here to measure the lensing cluster masses in Table~\ref{tbl:mass},
as well as the cluster masses $M_{200}$ and $M_{2500}$ in Table~\ref{tbl:props}.

\begin{deluxetable*}{lllllllll}
\tablewidth{0pt}
\tablecolumns{9}
\tablecaption{Comparisons between strong lensing and X-ray cluster mass
  estimates.}
\tablehead{
\colhead{Cluster} & 
\colhead{$\Delta$} &
\colhead{$r$} &
\colhead{$M_{\mathrm{lens}}(\leq r)$} &
\colhead{$M_{\mathrm{X-ray}}(\leq r)$} &
\colhead{$M_{\mathrm{lens}}(\leq r)/$} &
\colhead{Reduced} &
\colhead{Dynamical} &
\colhead{Reference} \\ & &
\colhead{$(h_{70}^{-1}$ Mpc)} &
\colhead{$(10^{14} \, h_{70}^{-1} \, M_\odot)$} &
\colhead{$(10^{14} \, h_{70}^{-1} \, M_\odot)$} & 
\colhead{$M_{\mathrm{X-ray}}(\leq r)$} &
\colhead{$\chi^2$} &
\colhead{State} & 
}
\startdata
ClG~2244$-$02 & 500$\, \Omega^{0.427}$ & $0.83^{+0.26}_{-0.20}$ &
$2.85^{+1.25}_{-0.99}$ &
$1.50^{+1.07}_{-0.63}$ & $1.90^{+2.81}_{-1.18}$ & 0.91 & Relaxed & 1 \\
& 18$\pi^2 \, \Omega^{0.427}$ & $1.31^{+0.42}_{-0.31}$ &
$4.22^{+1.63}_{-1.27}$ & $2.37^{+1.73}_{-0.99}$ &
$1.78^{+2.46}_{-1.06}$ &
0.86 & & 1 \\
Abell~370 & 500$\, \Omega^{0.427}$ & $1.15^{+0.28}_{-0.20}$ &
$6.45^{+2.04}_{-1.52}$ & $4.19^{+2.06}_{-1.30}$ &
$1.54^{+1.40}_{-0.75}$ & 0.85 & Unrelaxed & 1 \\
& 18$\pi^2 \, \Omega^{0.427}$ & $1.81^{+0.44}_{-0.32}$ &
$9.25^{+2.50}_{-1.92}$ &
$6.73^{+3.57}_{-2.16}$ & $1.37^{+1.20}_{-0.66}$  & 0.49 & & 1 \\
3C~220.1 & 500$\, \Omega^{0.427}$ & $1.17^{+0.45}_{-0.25}$ &
$3.22^{+1.37}_{-0.80}$ & $5.80^{+5.25}_{-2.19}$ &
$0.56^{+0.72}_{-0.34}$ & 0.44 & Relaxed & 1 \\
& 18$\pi^2 \, \Omega^{0.427}$ & $1.74^{+0.67}_{-0.37}$ &
$4.25^{+1.56}_{-0.92}$ &
$8.64^{+7.85}_{-3.27}$ & $0.49^{+0.59}_{-0.29}$ & 0.59 & &1 \\
MS~2137.3$-$2353 & 2500 & $0.46^{+0.02}_{-0.03}$ &
$1.62^{+0.18}_{-0.19}$ & $1.89^{+0.25}_{-0.31}$ &
$0.86^{+0.28}_{-0.19}$ & 0.65 & 
Relaxed & 2 \\
& 500$\, \Omega^{0.427}$ & $1.07^{+0.10}_{-0.06}$ &
$2.73^{+0.34}_{-0.30}$ & $3.16^{+0.60}_{-0.36}$ &
$0.86^{+0.23}_{-0.22}$ & 0.57 & & 1 \\
& 18$\pi^2 \, \Omega^{0.427}$ & $1.69^{+0.15}_{-0.10}$ &
$3.40^{+0.40}_{-0.37}$ & $4.99^{+0.95}_{-0.57}$ &
$0.68^{+0.18}_{-0.17}$ & 3.5 & & 1 \\
MS~0451.6$-$0305 & 500$\, \Omega^{0.427}$ & $1.38^{+0.25}_{-0.20}$ &
$13.4^{+3.1}_{-2.6}$ & $8.90^{+3.44}_{-2.31}$
& $1.50^{+1.00}_{-0.63}$ & 1.2 & Unrelaxed & 1 \\
& 18$\pi^2 \, \Omega^{0.427}$ & $2.09^{+0.38}_{-0.30}$ &
$18.3^{+3.7}_{-3.2}$ & $13.6^{+5.4}_{-3.6}$ &
$1.34^{+0.86}_{-0.55}$ & 0.68 & & 1 \\
MS~1137.5$+$6625 & 500$\, \Omega^{0.427}$ & $1.41^{+1.26}_{-0.45}$ &
$6.80^{+7.22}_{-2.64}$ & $12.5^{+32.0}_{-6.7}$
& $0.54^{+1.87}_{-0.45}$ & 0.082 & Relaxed & 1 \\
& 18$\pi^2 \, \Omega^{0.427}$ & $2.06^{+1.84}_{-0.66}$ &
$9.10^{+8.34}_{-3.08}$ & $18.2^{+46.9}_{-9.8}$
& $0.50^{+1.58}_{-0.41}$ & 0.099 & & 1 \\
Cl~0939$+$4713 & & $0.36$ & $0.38 \pm
0.05$ & $0.72 \pm 0.21$ & $0.53^{+0.31}_{-0.17}$ & 2.5 & Unrelaxed & 3 \\
& & $0.71$ & $0.69 \pm 0.08$ & $2.13 \pm 0.50$ &
$0.32^{+0.15}_{-0.09}$ & 8.1 & & 3 \\
ZwCl~0024$+$1652 & 500$\, \Omega^{0.427}$ & $0.94^{+0.39}_{-0.21}$ &
$2.02^{+0.97}_{-0.54}$ & $2.31^{+2.34}_{-0.91}$ &
$0.87^{+1.26}_{-0.56}$ & 0.027 & Unrelaxed & 1 \\
& 18$\pi^2 \, \Omega^{0.427}$ & $1.45^{+0.61}_{-0.32}$ &
$2.77^{+1.15}_{-0.64}$ & $3.59^{+3.63}_{-1.41}$ &
$0.77^{+1.03}_{-0.48}$ & 0.094 & & 1 \\
\enddata

\tablerefs{(1) \cite{OT04.2}; (2) \cite{AL01.2}; (3) \cite{DE03.1}.} 

\label{tbl:mass}
\end{deluxetable*} 

\subsection{Dynamical State of Clusters: \\ Relaxed vs. Unrelaxed}
\label{relax}

Since one of our aims is to measure the mass-temperature relation for
relaxed lensing clusters, we must determine which of the eight clusters in our sample
are dynamically relaxed.  
X-ray cluster mass estimates are based on the
assumption that the cluster is in hydrostatic equilibrium, and if a
cluster is relaxed it is also in hydrostatic equilibrium.  Therefore,
X-ray mass measurements for relaxed clusters should be accurate and
consistent with lensing mass measurements.

We use X-ray mass estimates from the literature, where the X-ray masses are measured for each cluster at two or three different radii.  For each cluster, Table~\ref{tbl:mass} gives the lensing mass and X-ray
mass measured 
within the two or three different cluster radii.  Table~\ref{tbl:mass} also shows the lensing mass to X-ray mass ratio and the
reduced $\chi^2$ of the comparison of lensing and X-ray masses.  For six
clusters, at all radii at which masses were measured, the ratio of lensing mass to X-ray mass is consistent
with unity and the reduced $\chi^2$ is $\lesssim 1$, suggesting
that these six clusters could be relaxed.  Additional observational evidence in \S~\ref{relaxed} and \S~\ref{unrelaxed} shows that four of these six
clusters are relaxed, while the
remaining two clusters are unrelaxed.

For at least one of the radii considered, the two clusters MS~2137$-$23 and Cl~0939$+$4713 each exhibit lensing to X-ray mass ratios that are
inconsistent with unity and reduced $\chi^2$ that are greater than
unity, which is evidence that the clusters are unrelaxed.  We measure
masses for MS~2137$-$23 within three different radii, and within one
of these radii the mass ratio is
inconsistent with unity and the reduced $\chi^2$ is greater than
unity.  However there 
is opposing evidence, given in \S~\ref{relaxed}, that characterizes MS~2137$-$23 as a
relaxed cluster. For Cl~0939$+$4713, the mass ratios measured at both
radii considered are inconsistent with unity and both reduced $\chi^2$ are much
greater than unity, suggesting Cl~0939$+$4713 may be an
unrelaxed cluster.  In \S~\ref{unrelaxed} we present more evidence in support of this
conclusion.  

Additional information about the dynamical state of a cluster can be found in its X-ray emission map.  For example, the position of the BCG relative to the peak in the cluster's X-ray profile may be evidence of a cluster's dynamical state: if the two are coincident the cluster is likely relaxed, otherwise it is likely unrelaxed.  The centroid shift is one means of quantifying this positional difference (e.g., \citealt{MO93.1, JE08.1}).  Additionally, a smooth distribution of X-ray gas indicates the cluster is likely in a relaxed state.  However, if the X-ray gas is distributed irregularly or shows evidence of shocks or substructure, the cluster is likely unrelaxed and undergoing a merger. 
Below we examine evidence for the dynamical state of each cluster
individually and label 
each cluster as relaxed or unrelaxed (these labels are also given in
Table~\ref{tbl:mass}).  We first discuss the four relaxed clusters, then the
four unrelaxed clusters. 

\subsubsection{Relaxed Clusters}
\label{relaxed}

\begin{itemize}

\item \textbf{Cl~2244$-$02}:
We find that X-ray and lensing masses for Cl~2244$-$02 are consistent
(Table~\ref{tbl:mass}) and \cite{OT98.1} also find 
consistent X-ray and lensing masses, suggesting that hydrostatic
equilibrium is a valid assumption for Cl~2244$-$02 and that it is a relaxed
cluster. 

\item \textbf{3C~220.1}: The radial profile of X-ray emission from
  3C~220.1 shows no sign of
irregularity and the profile is well-fit by a model assuming
hydrostatic equilibrium, which suggest that 3C~220.1 is a relaxed
cluster \citep{WO01.1}.

\item \textbf{MS~2137$-$23}: The X-ray and strong lensing masses of MS~2137$-$23 are in good agreement \citep{AL98.1}, indicating that it is in a relaxed state.  Many relaxed clusters also have cooling flows, such as the massive cooling flow in MS~2137$-$23
\citep{AL98.1, WU00.1}.

\item \textbf{MS~1137$+$66}: The cluster MS~1137$+$66 not only has consistent X-ray and weak lensing
masses (Table~\ref{tbl:mass}), but also has a small centroid shift \citep{MA08.2} and may host a moderate cooling flow \citep{DO99.1}.  In
addition, Sunyaev Zel'dovich observations of the cluster show no
obvious substructure \citep{CO02.1}.  These properites connote that
MS~1137$+$66 is a relaxed cluster. 

\begin{deluxetable*}{llllll}
\tablewidth{0pt}
\tablecolumns{6}
\tablecaption{Cluster lensing masses and
  X-ray temperatures.} 
\tablehead{
\colhead{Cluster} & 
\colhead{$z$} &
\colhead{$M_{200}$} & 
\colhead{$M_{2500}$} & 
\colhead{$k T$} &
\colhead{Reference} \\ &&
\colhead{$(10^{14} \, h_{70}^{-1} \, M_\odot)$} &
\colhead{$(10^{14} \, h_{70}^{-1} \, M_\odot)$} &
\colhead{(keV)} 
}
\startdata
ClG~2244$-$02 & 0.33 & $4.5 \pm 0.9$ & $1.3 \pm 0.2$ & $4.85^{+1.25}_{-0.96}$ & 1 \\
Abell~370 & 0.375 & $9.0 \pm 1.0$ & $2.9 \pm 0.3$ & $7.20^{+0.75}_{-0.77}$ & 1 \\
3C~220.1 & 0.62 & $3.1 \pm 0.3$ & $0.91 \pm 0.10$ & $5.6^{+1.5}_{-1.1}$ & 2 \\
MS~2137.3$-$2353 & 0.313 & $2.9 \pm 0.4$ & $1.5 \pm 0.2$ & $4.57^{+0.41}_{-0.35}$ & 1 \\
MS~0451.6$-$0305 & 0.55 & $18 \pm 2$ & $6.3 \pm 0.7$ & 
$8.62^{+1.54}_{-1.21}$ & 1 \\
MS~1137.5$+$6625 & 0.783 & $6.5 \pm 0.7$ & $1.5 \pm 0.2$ & $6.70^{+1.84}_{-1.46}$ & 1 \\
Cl~0939$+$4713 & 0.41 & $0.71 \pm 0.11$ & $0.21 \pm 0.03$ & $7.6^{+2.8}_{-1.6}$ & 3 \\
ZwCl~0024$+$1652 & 0.395 & $2.3 \pm 0.2$ & $0.69 \pm 0.07$ & $5.17^{+1.95}_{-1.34}$
& 1 \\
\enddata

\tablerefs{(1) \cite{HO01.1}; (2) \cite{OT00.1}; (3) \cite{SC98.3}.}
 
\label{tbl:props}
\end{deluxetable*} 

\subsubsection{Unrelaxed Clusters}
\label{unrelaxed}

\item \textbf{Abell~370}: Abell~370 hosts two cD galaxies, and there are X-ray peaks centered on each cD
\citep{ME94.1}.  The two 
cD galaxies are moving relative to each other at 1000 km s$^{-1}$,
signaling that Abell~370 is an unrelaxed cluster undergoing a merger
\citep{KN93.1}. 

\item \textbf{Cl~0939$+$4713}: X-ray observations of Cl~0939$+$4713 show evidence for substructure
\citep{SC96.3}, and the disagreement between lensing and X-ray masses
shown in Table~\ref{tbl:mass} further suggests that Cl~0939$+$4713 is not in hydrostatic
equilibrium.  These observations indicate Cl~0939$+$4713 is an unrelaxed cluster.  

\item \textbf{Cl~0024$+$17}: The two dark matter clumps near the center of Cl~0024$+$17 are separated in redshift, implying that it is a merging cluster \citep{NA09.1}. There is additional evidence for substructure in Cl~0024$+$17 in its mass
models, which require substructure to produce a good fit to the cluster's
lensing arcs \citep{BR00.1}.  The redshifts of the member galaxies are
distributed bimodally, fortifying the evidence that 
Cl~0024$+$17 may have undergone a merger with another cluster
\citep{CZ02.1}.  The evidence implies that Cl~0024$+$17 is an unrelaxed cluster.

\item \textbf{MS~0451.6$-$0305}: The distribution of mass within the central 1$^\prime$ of MS~0451.6$-$0305 is not smooth, and the centroid shift indicates the BCG is not located at the X-ray peak \citep{BO04.2, MA08.2}.  These observations suggest that MS~0451.6$-$0305 is unrelaxed.

\end{itemize}

\subsection{Cluster X-ray Temperatures}
\label{temp}

The temperature of the intracluster medium is commonly measured
using its X-ray emission in one of several ways:  
through fits to the cluster's observed X-ray spectrum (yielding the spectroscopic temperature $T_{\rm s}$), through
weighting by the mass of the gas element (yielding the mass-weighted temperature
$T_{\rm m}$), or through weighting by the emissivity of the gas
element (yielding the emission-weighted temperature $T_{\rm em}$).  However, the spectroscopic temperature $T_{\rm s}$
is systematically lower than the mass-weighted temperature $T_{\rm m}$ and
the emission-weighted temperature $T_{\rm em}$
\citep{MA01.2, MA04.1}, so an accurate temperature comparison across different clusters requires consistent temperature measurements.
 
To ensure that the cluster temperatures we use for our sample are as
consistent as possible, we use the mean cluster temperatures derived
from the single-temperature model fits of {\it ASCA} data in \cite{HO01.1}.  This large,
homogeneous catalog of spectroscopic cluster temperatures 
includes six of our clusters, and for the
remaining two clusters, 3C~220.1 and Cl~0939$+$4713, we remain as
consistent as possible by using
spectroscopic temperatures from single-temperature
fits.   Table~\ref{tbl:props} gives the cluster
temperatures and the corresponding references. We note that none
of the temperatures we use apply corrections for 
cool cores at the cluster centers.

Some clusters in our sample also have temperature measurements from {\it Chandra} and {\it XMM-Newton} data.  Specifically, 3C~220.1 has a {\it Chandra} temperature of $8.5^{+3.7}_{-2.3}$ keV within 10 -- 45$^{\prime\prime}$ \citep{WO01.1}; MS~0451.6$-$0305 has a {\it Chandra} temperature of $6.7^{+0.6}_{-0.5}$ keV within $r_{500}$ \citep{MA08.2}; MS~1137.5$+$6625 has a {\it Chandra} temperature of $5.8^{+0.7}_{-0.6}$ keV within $r_{500}$ \citep{MA08.2}; and Cl~0024$+$17 has an average {\it Chandra} temperature of $4.47^{+0.83}_{-0.54}$ keV  \citep{OT04.1} and an {\it XMM-Newton} temperature of $3.52 \pm 0.17$ keV within 3$^\prime$ \citep{ZH05.2}.  Use of {\it Chandra} or {\it XMM-Newton} temperatures could change the results of the mass-temperature relation.  However, because {\it Chandra} and {\it XMM-Newton} temperatures have been measured for only a subset of our sample, and because these temperatures are measured within inconsistent cluster radii, we do not use {\it Chandra} and {\it XMM-Newton} measurements in our determination of the cluster mass-temperature relation below.

\section{The Mass-Temperature Relation}
\label{mtrelation}
Theoretical arguments suggest a correlation between cluster mass and
X-ray temperature for relaxed clusters, which provides the link
between the gas in a cluster and its mass.  Here we determine the
cluster mass-temperature relation for relaxed {\it strong lensing} clusters, and we also explore the
correlation between the scatter in cluster temperature and the scatter
in cluster concentration to establish a general mass-temperature relation that is independent of the dynamical state of the clusters. 

\subsection{The $M-T$ Relation for Relaxed Strong Lensing Clusters}
\label{mtrelaxed}

A correlation between cluster mass and cluster X-ray gas temperature in relaxed clusters
is expected as a direct consequence of theoretical arguments.  
If a cluster's X-ray gas is in virial and hydrostatic equilibrium,
then the theoretical expectation is that cluster mass scales with X-ray temperature as
$E(z)M_\Delta=A(\Delta)T^{1.5}$, where
$E(z)=H(z)/H_0=\sqrt{\Omega_{\rm m}(1+z)^3 + \Omega_\Lambda}$ for a flat
Universe, $M_\Delta$ is the cluster mass within the radius where
the mean mass density is $\Delta$ times the critical density, and $A(\Delta)$ is the $\Delta$-dependent normalization.  
  
The critical overdensity $\Delta=2500$ is commonly used in cluster
analyses because in the central regions enclosed by $r_{2500}$, {\it Chandra} 
cluster temperature profiles can be measured even at high redshifts
(e.g., up to $z=0.9$ in \citealt{AL04.2}).  The overdensity
$\Delta=2500$ is therefore appropriate for our cluster sample, which
extends to $z=0.8$.  Using the overdensity $\Delta=2500$, we can write
the cluster mass-temperature relation in power law form as

\begin{equation}
E(z) \left( \frac{M_{2500}} {10^{14} \, h_{70}^{-1} \, M_\odot} \right) = A \left( \frac{k T}
{5 \, \mathrm{keV}} \right)^\alpha \, .
\end{equation} 

\begin{figure}[!t]
\begin{center}
\includegraphics[scale=.9]{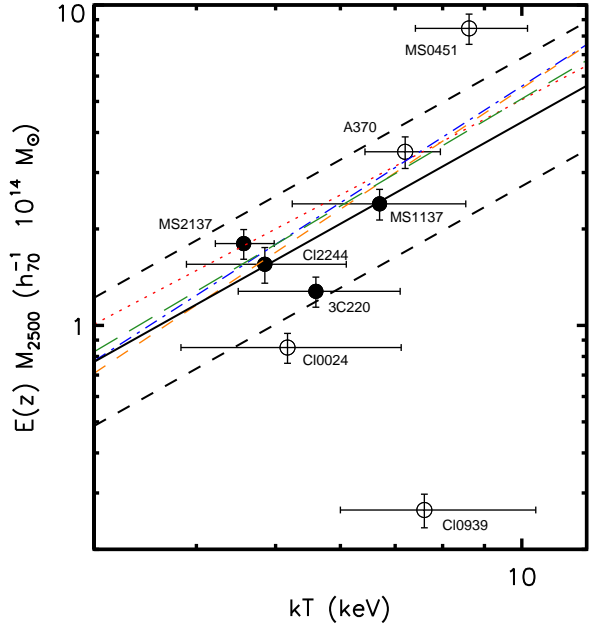}
\end{center}
\vspace{-.1in}
\caption{The mass-temperature relation for observed strong lensing clusters.  Unrelaxed clusters (open circles) are not included in the fit, and the relaxed clusters (black points) are fit by a power law with slope $\alpha=1.43$ (black solid line).  The 1$\sigma$ scatter for all eight clusters is large, $\Delta(\log[E(z) M_{2500}])=0.2$ (black dashed lines). Also shown are the other $M-T$ relations for observational samples that use spectroscopic temperatures as we do: 17 weak lensing clusters with $3.6 < T_{\rm s}  \; ({\rm keV}) < 9.8$ (\citealt{HO07.2}; red dotted line), 13 relaxed X-ray clusters with $0.7 < T_{\rm s}  \; ({\rm keV}) < 8.9$ (\citealt{VI06.1}; blue dash-dotted line), 10 relaxed X-ray clusters with $2.2 < T_{\rm s}  \; ({\rm keV}) < 8.3$ (\citealt{AR05.1}; orange dashed line), and six relaxed X-ray clusters with $3.7 < T_{\rm s}  \; ({\rm keV}) < 8.3$ (\citealt{AR05.1}; green long dashed line).  We find that our slope is in agreement with both the theoretical expectation of $\alpha=1.5$ and measurements of $\alpha$ by other observations. For a detailed comparison to these and other estimates of the $M-T$ relation, see Table~\ref{tbl:mtrelation}.       
}
\label{fig:MT}
\end{figure}

Using our sample of four dynamically relaxed lensing clusters given in
\S~\ref{relaxed}, a best fit to the power law $M-T$ relation yields 
$A=1.60 \pm 3.42$ and $\alpha=1.43 \pm 1.28$, consistent with the theoretical expectation of
$\alpha=1.5$. Figure~\ref{fig:MT} shows this best-fit relation, for which the RMS scatter is 360$\%$ for all
eight clusters and 500$\%$ for the four unrelaxed clusters.

\begin{deluxetable*}{llllll}
\tablewidth{0pt}
\tablecolumns{6}
\tablecaption{Power Law Fits to the $M-T$ Relation.}
\tablehead{
\colhead{$A$} & 
\colhead{$\alpha$} &
\colhead{Method\tablenotemark{a}} &
\colhead{$k T$ (keV)\tablenotemark{b}} &
\colhead{Sample} & 
\colhead{Reference}  
}
\startdata
$1.60 \pm 3.42$ & $1.43 \pm 1.28$ & SL & $4.6 < T_{\rm s} < 6.7$
& 4 relaxed SL clusters & 1 \\
$2.0\phantom{0} \pm 0.29$ & $1.34^{+0.30}_{-0.28}$ & WL & $3.6 < T_{\rm s} < 9.8$ & 17 WL
clusters & 2 \\ 
$1.79 \pm 0.07$ & $1.64 \pm 0.06$ & X-ray & $0.7 < T_{\rm s} < 8.9$ & 13 relaxed
clusters & 3 \\
$2.06 \pm 0.10$ & $1.58 \pm 0.07$ & X-ray & $0.6 < T_{\rm m} < 9.3$ & 13 relaxed
clusters & 3 \\
$1.69 \pm 0.05$ & $1.70 \pm 0.07$ & X-ray & $2.2 < T_{\rm s} < 8.3$ & 10
relaxed clusters & 4 \\
$1.79 \pm 0.06$ & $1.51 \pm 0.11$ & X-ray & $3.7 < T_{\rm s} < 8.3$ & 6
relaxed clusters & 4 \\
$1.88 \pm 0.26$ & $1.52 \pm 0.36$ & X-ray & $5.6 < T_{\rm m} < 15.3$ & 5
relaxed WL or SL clusters & 5 \\
$1.97 \pm 0.07$ & $1.54 \pm 0.02$ & Simulation & $T_{\rm m}$ & $M_{2500} > 4 \times
10^{14} \, h_{70}^{-1} \, M_\odot$ clusters & 6 \\ 
 & & & &  in hydrodynamics
simulation & \\
\enddata

\tablenotetext{a}{Method used to determine the cluster mass, where SL is strong lensing and WL is weak lensing.}
\tablenotetext{b}{Temperature range of the cluster sample, where $T_{\rm s}$
is the spectroscopic temperature and $T_{\rm m}$ is the mass-weighted temperature.}

\tablerefs{(1) This paper; (2) \cite{HO07.2}; (3) \cite{VI06.1}; (4) \cite{AR05.1};
  (5) \cite{AL01.2}; (6) \cite{KA05.1}.} 

\label{tbl:mtrelation}
\end{deluxetable*} 

\begin{figure}[!t]
\begin{center}
\includegraphics[scale=.9]{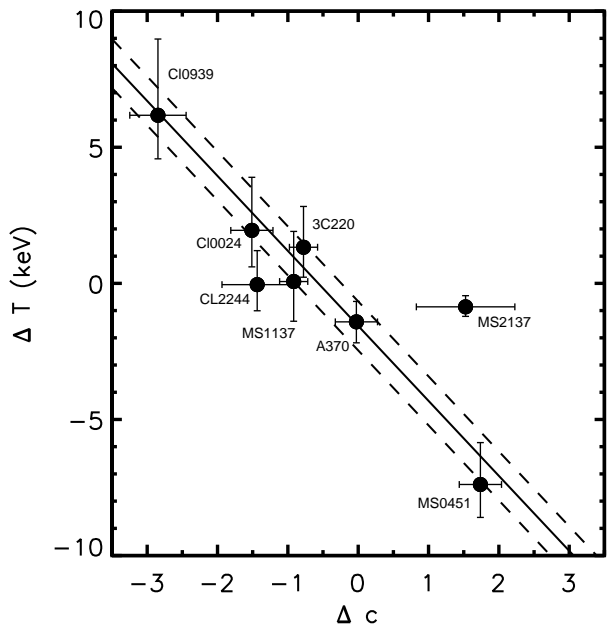}
\end{center}
\vspace{-.1in}
\caption{The correlation between the difference $\Delta T$ between the observed X-ray temperatures and the predicted temperatures from the $M-T$ relation and the difference $\Delta c$ between the measured concentrations and the predicted concentrations from the $c-M$ relation.  The eight strong lensing clusters in our sample are represented, and the solid line shows the best-fit line to the data $\Delta T = (-2.75 \; {\rm keV}) \Delta c -(1.56 \; {\rm keV})$.  The dashed lines show the 1$\sigma$ scatter $\Delta (\Delta T)=0.9$ keV.
}
\label{fig:deltaTdeltac}
\end{figure}

\begin{figure}[!t]
\begin{center}
\vspace{-.1in}
\includegraphics[scale=.9]{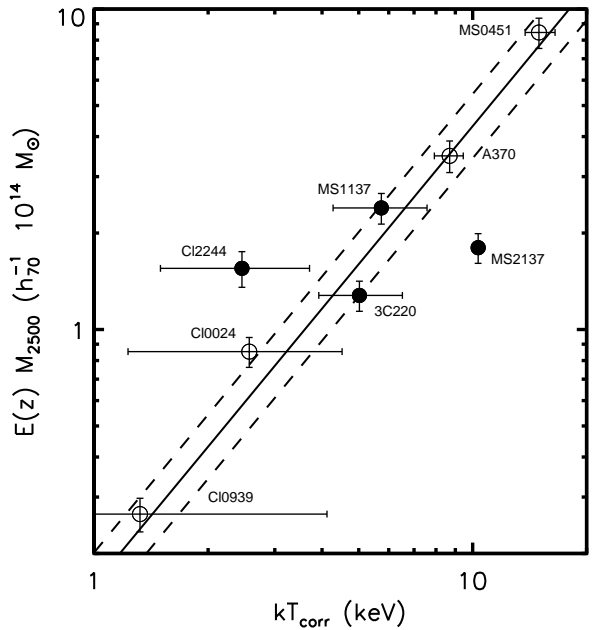}
\end{center}
\caption{The mass-temperature relation, after correcting for the scatter in temperature, for observed strong lensing clusters.  As in Figure~\ref{fig:MT}, open circles represent unrelaxed clusters and black points represent relaxed clusters.  We adjust the temperature of each cluster according to its concentration and the $\Delta T - \Delta c$ relation.  The best-fit $M-T$ relation for relaxed clusters, derived in \S~\ref{mtrelaxed}, is shown as the solid line. The 1$\sigma$ scatter for all eight clusters is $\Delta(\log[E(z) M_{2500}])=0.1$ (black dashed lines), significantly smaller than the scatter in the uncorrected $M-T$ relation (see Figure~\ref{fig:MT}). 
}
\label{fig:mtscatter}
\end{figure}

We compare with other observations and simulations of the $M-T$ relation in
Table~\ref{tbl:mtrelation}, including those that used spectroscopic
temperatures $T_{\rm s}$ and those that used mass-weighted
temperatures $T_{\rm m}$.  
For cases where the
temperature normalization is not 5 keV and/or the mass scaling is not $10^{14} \, h_{70}^{-1} \, M_\odot$, we recalculate $A$ using the
published slope $\alpha$, a temperature normalization of 5 keV, and a
mass normalization of $10^{14} \, h_{70}^{-1} \, M_\odot$.  To
be conservative, we assume the fractional error in $A$ is unchanged.

The observations we compare span varying temperature ranges, and
there is some evidence that the $M-T$ relation steepens for cooler
clusters (e.g., \citealt{NE00.2,FI01.1}); for example, \cite{AR05.1} find a slope of $\alpha=1.51$ for clusters with $3.7 < T_{\rm s} < 8.3$ keV, which increases to $\alpha=1.70$ for clusters with $2.2 \; \mathrm{keV} < T_{\rm s} < 8.3$ keV.  The temperature range we probe ($4.6 \; \mathrm{keV} < T_{\rm s} < 6.7 $ keV) is likely too small to exhibit a significant change in slope, but we lack a large enough statistical sample to test this properly.

We find that our best-fit slope $\alpha$ is consistent with both
the theoretical expectation and the slopes derived by other
observations and simulations of clusters.  Our best-fit normalization
$A$ is somewhat lower than, but still consistent with, the
normalizations found by the other observations and
simulations. We find that relaxed strong lensing clusters follow the same
$M-T$ relation as relaxed clusters in general.

\subsection{Correlation between the Temperature Scatter and Concentration Scatter}
\label{dTdc}

We have derived an $M-T$ power-law relation for relaxed lensing clusters, but a more general $M-T$ relation including both relaxed and unrelaxed clusters may be possible if we account for the differences in cluster concentrations.  First, we define the virial radius of a cluster as the radius $r_\mathrm{vir}$ at which the average cluster density equals $\Delta_\mathrm{vir}(z)$ times the mean density at the cluster redshift $z$, where $\Delta_\mathrm{vir}(z) \simeq (18 \pi^2 + 82x -39x^2)/(1+x)$ and $x  \equiv \Omega_{\rm m}(z) -1$ \citep{BR98.2}.  Using the scale radius $r_{\rm s}$ of the best-fit NFW profile to each cluster, the cluster concentration is defined as $c_\mathrm{vir} \equiv r_\mathrm{vir} / r_{\rm s}$.

Since more concentrated clusters are expected to form at higher redshifts
(e.g., \citealt{NA97.1}; \citealt{WE02.1}), if the cluster X-ray gas
cools with time there might be a correlation between high cluster
concentrations and low cluster temperatures.  In addition, mergers
with other clusters or groups may deplete the central mass densities in clusters while shock-heating the cluster gas, producing high cluster temperatures for low cluster concentrations.  Here, we analyze whether there is any such correlation between the scatter in temperature and the scatter in concentration for our sample of eight strong lensing clusters.    

Cluster concentrations $c_\mathrm{vir}$ and cluster virial masses $M_\mathrm{vir} \equiv M(\leq r_\mathrm{vir})$ are determined by strong lensing measurements for each of the clusters in our sample in \cite{CO07.1}.  
The concentration $c_\mathrm{vir}=16$ determined by strong lensing measurements of MS~2137.3$-$2353 is known to be overestimated  because the cluster's dark matter halo is likely elongated along or near the line of sight \citep{GA05.1}, so we instead use the concentration $c_\mathrm{vir}=8.75$ derived from the X-ray mass profile for MS~2137.3$-$2353 \citep{SC07.2}.  We note that if the lensing concentration were used for MS~2137.3$-$2353, Equation~\ref{eq:dtdc} would be $\Delta T = (-0.07 \; {\rm keV}) \Delta c -(0.49 \; {\rm keV})$.

From a sample of 62 galaxy clusters, \cite{CO07.1} find a power-law relation between cluster concentration $c_\mathrm{vir}$ and cluster virial mass $M_\mathrm{vir}$ of 
\begin{equation}
c_\mathrm{vir}=\frac{14.5 \pm 6.4} {(1+z)} \left( \frac{M_\mathrm{vir}} {1.3 \times 10^{13} \, h^{-1} \, M_\odot} \right)^{-0.15 \pm 0.13} \; ,
\end{equation}
where $z$ is the cluster redshift.  For each of the eight clusters in our sample, we calculate the difference $\Delta c$ between the measured concentration and the concentration predicted by the above $c-M$ relation.  We also calculate the difference $\Delta T$ between the measured X-ray temperature and the temperature predicted by the $M-T$ relation we determined in \S~\ref{mtrelaxed} for the four relaxed clusters.

Figure~\ref{fig:deltaTdeltac} shows the results of these $\Delta T$
and $\Delta c$ calculations.  The best-fit line to the data is 
\begin{equation}
\Delta T = (-2.75 \; {\rm keV}  \pm 0.07 \; {\rm keV}) \Delta c -(1.56 \; {\rm keV} \pm 0.49 \; {\rm keV}) \; , 
\label{eq:dtdc}
\end{equation}
suggesting that indeed higher (lower) temperature clusters tend to have lower (higher) concentrations.

\subsection{The $M-T$ Relation for All Strong Lensing Clusters}

Using the relation between the scatter in cluster temperature
and the scatter in cluster concentration for the eight strong lensing
clusters (\S~\ref{dTdc}), we adjust for the apparent dependence of cluster temperatures on
cluster concentrations.  We use $\Delta c$ for each cluster to calculate its
corresponding $\Delta T$ from the best-fit relation given in Equation~\ref{eq:dtdc}.  We then subtract this $\Delta
T$ from the measured temperature to obtain a corrected temperature $T_{\rm corr}$, and we illustrate the resultant
temperature-corrected $M-T$ relation in Figure~\ref{fig:mtscatter}.
The figure also shows the relation we derived in \S~\ref{mtrelaxed} for the four
relaxed clusters, where $A=1.60$ and $\alpha=1.43$.

We find that cluster concentration, mass, and X-ray temperature are
tightly correlated, and as a result incorporating the $\Delta T - \Delta c$
relation significantly reduces the scatter in the $M-T$ relation.
Comparing Figure~\ref{fig:mtscatter} to Figure~\ref{fig:MT}
underscores the impact of our temperature correction in reducing the
scatter in the $M-T$ relation.  The temperature correction reduces the
RMS scatter for all eight clusters by a factor of 6, from 360$\%$ to
60$\%$, and more significantly, reduces the RMS scatter for the four unrelaxed clusters by a factor of 30, from 500$\%$ to 15$\%$. (The RMS scatter for the four relaxed clusters increases from 26$\%$ to 83$\%$, possibly because the temperatures we use do not correct for cool cores at the cluster centers.)
With the temperature correction, even unrelaxed clusters follow the
$M-T$ relation we originally derived using only the relaxed clusters
(\S~\ref{mtrelaxed}).  Therefore, we suggest this temperature
correction as a tool for establishing a universal $M-T$ relation that
applies to all galaxy clusters regardless of their dynamical state.  

The error in the measurement of $\sigma_8$ from cluster counts depends
directly on the error in the cluster $M-T$ relation; for example, a
25$\%$ 1$\sigma$ uncertainty in the zero point of the $M-T$ relation corresponds
to a $10\%$ 1$\sigma$ uncertainty in $\sigma_8$ \citep{EV02.1}.  Consequently,
 we find that the temperature correction not only reduces the
scatter in the $M-T$ relation, but also significantly reduces the error in the
corresponding measurement of $\sigma_8$.

An alternate cluster scaling relation that also has lower scatter than the traditional $M-T$ relation is the $Y_{\rm X} - M_{500}$ relation \citep{KR06.1}.  
Here, $M_{500}$ is the cluster mass within the radius $r_{500}$ enclosing an overdensity of 500 relative to the critical density, $Y_{\rm X}=M_g T_{\rm X}$, $M_g$ is the cluster gas mass within $r_{500}$, and $T_{\rm X}$ is the mean spectral X-ray temperature of the cluster.  However, this scaling relation is limited by the assumptions that the gas is both spherically distributed and in hydrostatic equilibrium.  Our scaling relation offers the advantage that it is based on lensing mass estimates that are free of these assumptions.  

\begin{deluxetable*}{lllll}
\tablewidth{0pt}
\tablecolumns{5}
\tablecaption{BCG luminosities.} 
\tablehead{
\colhead{Cluster} & 
\colhead{BCG\tablenotemark{a}} &
\colhead{$L_\mathrm{K, BCG}$} & 
\colhead{$L_\mathrm{K, passive, BCG}$} & 
\colhead{Reference} \\ & &
\colhead{$(10^{11} \, h_{70}^{-2} \, L_\odot)$} &
\colhead{$(10^{11} \, h_{70}^{-2} \, L_\odot)$} & 
}
\startdata
ClG~2244$-$02 & & $1.03 \pm 0.09$ & $0.96 \pm 0.09$ &  1 \\
Abell~370 & G1 & $1.5\phantom{0} \pm 0.1$ & $1.3\phantom{0} \pm 0.1$ &  \\ 
3C~220.1 &  & $6.4\phantom{0} \pm 0.4$ & $5.3\phantom{0} \pm 0.3$ &  \\
MS~2137.3$-$2353 & & $8.98 \pm 0.09$ & $0.84 \pm 0.08$ &  \\
MS~0451.6$-$0305 & & $4.3\phantom{0} \pm 0.3$ & $3.7\phantom{0} \pm 0.3$ &  2 \\
MS~1137.5$+$6625 & & $15\phantom{.0} \pm 2$ & $11\phantom{.0} \pm 1$ &  \\
Cl~0939$+$4713 & G1 & $1.9\phantom{0} \pm 0.2$ & $1.7\phantom{0} \pm 0.2$ &  3 \\ 
ZwCl~0024$+$1652 & $\#$362 & $1.69 \pm 0.07$ & $1.51 \pm 0.06$ &  4, 5 \\ 
\enddata

\tablenotetext{a}{See \cite{CO06.1} for identification of the galaxies
by name.}

\tablerefs{(1) \cite{BA82.2}; (2) \cite{EL98.1}; (3) \cite{DE03.1}; (4) \cite{KN03.1};
  (5) \cite{MO05.1}.} 

\label{tbl:lum}
\end{deluxetable*} 

\section{BCG Properties}
\label{bcgprop}

In addition to the interdependencies of many cluster properties, properties of the BCG have also been shown to correlate with the host cluster.  Here we identify the BCG in each of our clusters, measure the luminosity of each BCG, and examine the correlation between BCG luminosity and host cluster mass for our strong lensing sample.

\subsection{BCG Determination}
\label{bcg}

We select each cluster's BCG as the brightest member galaxy.  Each BCG corresponds to the lens galaxy or one of the lens galaxies used to determine the cluster mass distribution in \cite{CO06.1}.  When multiple lens galaxies were used to model a single cluster, we identify which of the lens galaxies is the BCG in Table~\ref{tbl:lum}, and we also note references that confirm the BCG selection.

\subsection{BCG Luminosity Determination}
\label{light}

For each cluster we have {\it HST} imaging taken in some combination of the
filters F450W, F555W, F675W, F702W, and F814W. Using Source EXtractor
\citep{BE96.1}, we measure MAG\_AUTO magnitudes for the BCG galaxies.
We estimate the 
magnitude uncertainties by adding in quadrature the error in the measured flux
and the estimated background subtraction error, which is the product of the
area of the extraction aperture and the RMS variation of the subtracted
background flux.  We calculate the BCG luminosities using the available
photometry in an observed band as the normalization factor on two types of
spectral energy distribution templates, and then compute the rest-frame
magnitudes and luminosities in several bands including {\it K}-band.  The templates we
use are calculated from the
\cite{BR03.1} stellar population synthesis models with a Salpeter
initial mass function.  The first we use is a fixed-age 10 Gyr old simple
stellar population, and the second is for a simple stellar population with
an age given by an assumed formation redshift of $z=3.0$. The latter
enables an estimate of the passively-evolved BCG luminosity.

\section{The BCG Luminosity - Cluster Mass Relation}
\label{lm}

\begin{figure}[!t]
\begin{center}
\includegraphics[scale=.9]{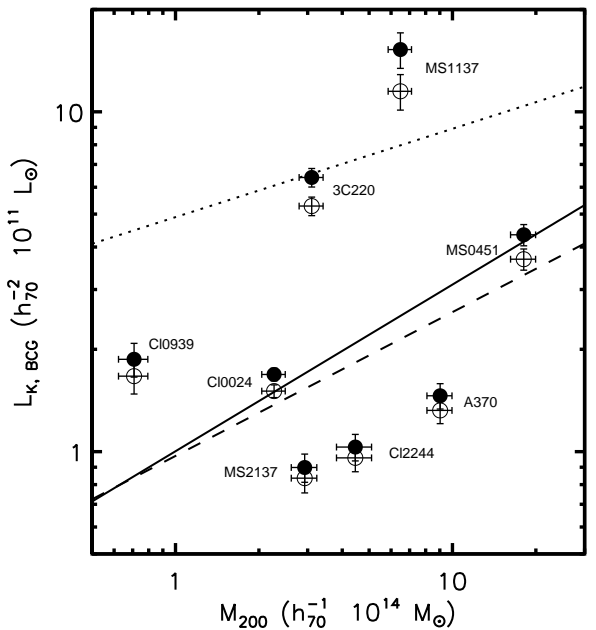}
\end{center}
\vspace{-.1in}
\caption{The correlation between {\it K}-band BCG luminosity and cluster mass for our sample of strong lensing clusters.  Uncorrected luminosities (black points) are fit by the solid line, while luminosities corrected for passive evolution (open circles) are fit by the dashed line.  For comparison, the \cite{LI04.3} $L-M$ relation for the general cluster population is shown as the dotted line.  Our best-fit power laws are significantly steeper than that of \cite{LI04.3}, hinting that BCGs in lensing clusters may have different formation histories than BCGs in typical clusters. }
\label{fig:lmlinmohr}
\end{figure}

Although it is still unclear how BCGs form, conventional formation scenarios include galactic
cannibalism, cooling flows, and mergers during cluster formation (\S~\ref{intro}).
The evolution of the luminosity of the BCG with the mass of the cluster may distinguish between these models and offer insight into the formation of BCGs.
Semianalytic and numerical simulations of structure formation suggest
a tight correlation between BCG luminosities and cluster masses (e.g.,
\citealt{SO99.1,CO00.1}), and we can parameterize such a correlation
between {\it K}-band BCG luminosities and cluster masses $M_{200}$ by the power law

\begin{equation}
\frac{L_\mathrm{K, BCG}} {10^{11} \; h_{70}^{-2} \; L_\odot} = B \left( \frac{M_{200}} {10^{14} \; h_{70}^{-1} \; M_\odot} \right)^\beta \; .
\end{equation}

Here, we examine the relation between BCG luminosity and cluster mass for clues about the
formation histories of BCGs in strong lensing clusters and how their
formations may differ from the general BCG population.  We represent
the general BCG population with the \cite{LI04.3} study of 93 BCGs at $z \leq 0.09$ in the Two Micron All Sky Survey (2MASS).

For an accurate comparison to the $L-M$ relation 
\cite{LI04.3} find from 2MASS, we follow their definition of BCG
luminosity.  \cite{LI04.3} measure BCG luminosities in the {\it K}-band using 20 mag arcsec$^{-2}$ isophotal elliptical
aperture magnitudes for 2MASS, called K20 magnitudes.  Similarly, we
convert to the {\it K}-band (see  \S~\ref{light}) and measure BCG
magnitudes using SExtractor's MAG\_AUTO function \citep{BE96.1}, which
has good agreement with 2MASS K20 total magnitudes for sources such as
BCGs that are bright and extended \citep{EL06.1}.  We then convert the
magnitudes into {\it K}-band luminosities as described in
\S~\ref{light}. The resultant {\it K}-band BCG luminosities, along
with the luminosities corrected for passive evolution, are given in Table~\ref{tbl:lum}.

Figure~\ref{fig:lmlinmohr} illustrates the correlation between BCG luminosities and
cluster masses $M_{200}$. We find
the best-fit power law to the data is given by $B=0.97 \pm
0.17$ and $\beta=0.48 \pm 0.09$ for all strong lensing clusters (solid
line in Figure~\ref{fig:lmlinmohr}) and $B=0.93 \pm 0.18$ and $\beta=0.39 \pm 0.10$ for all strong
lensing clusters when the BCG luminosities are corrected for passive
evolution (dashed line in Figure~\ref{fig:lmlinmohr}).   The similarity of these two results implies that the passive evolution of BCG
luminosities with redshift has little effect on the $L-M$ relation,
and more generally
there is no evidence for evolution in the $L-M$ relation from
$z\sim1$ to $z\sim0$ \citep{BR08.1}

For comparison, \cite{LI04.3} find a
best-fit power law of $B=4.9 \pm 0.2$ and $\beta=0.26 \pm 0.04$
(dotted line in Figure~\ref{fig:lmlinmohr}), which is consistent with the slopes found by
analytic estimates and cosmological simulations of the growth of central galaxies.  Using the
galaxy-dark matter correlation function to determine host dark matter
halo masses for observational catalogs of galaxies, \cite{CO05.1} find $L
\propto M_{200}^{<0.3}$ for halo masses $\gtrsim 4 \times
10^{13} \; h^{-1} \; M_\odot$.  Similarly, \cite{VA06.1} determine a
correlation of $L
\propto M_{100}^{0.28}$ when they combine the subhalo mass distribution
derived from simulations with an empirical galaxy luminosity function.
They also find little dependence of the $L-M$ relation on waveband. 

From their slope of $\beta=0.26$, \cite{LI04.3} conclude that while
other cluster members may merge with BCGs and increase BCG
luminosities, such effects are not sufficient to fully account for the
growth in $L_\mathrm{K, BCG}$ with cluster mass.  Instead,
\cite{LI04.3} suggest that BCGs must grow mainly through mergers with
other BCGs brought in when the host galaxy cluster merges with other
groups or clusters.  In addition to the many hierarchical structure
formation simulations and models that support this scenario (e.g.,
\citealt{ME85.1, DU98.1, BO06.1}), there are also observations of a
pair of $\sim L^{*}$ elliptical galaxies merging to build up the BCG
in a rich cluster at $z=1.26$ \citep{YA02.1}.

Our slope $\beta$ is 50$\%$ (when luminosities are corrected
for passive evolution) to 85$\%$ (when luminosities are not corrected
for passive evolution) steeper than that of \cite{LI04.3}, hinting
that strong lensing clusters may undergo more mergers with groups and
clusters, or merge with more massive groups and clusters, than the
average cluster.  Both more mergers and mergers with more
massive systems could account for the initial evidence for an increase in
$L_\mathrm{K, BCG}$ with cluster mass we find in strong lensing
clusters, and would also be consistent with simulations that suggest
strong lensing clusters are dynamically more active than the general
cluster population \citep{BA96.2}.  However, the scatter in our $L-M$ relation is significant, and a larger sample of strong lensing clusters is necessary to draw definitive conclusions about the formation histories of strong lensing clusters.

\section{Conclusions}
\label{conclude}

We have determined the scaling of cluster mass with cluster temperature and the scaling of BCG
luminosity with cluster mass for eight observed strong lensing
galaxy clusters imaged with {\it HST} and at redshifts $0.3 < z < 0.8$.  We explored cluster
concentrations as a means of reducing the scatter in the $M-T$
relation and
enabling more precise constraints on cosmological parameters, and we
used the $L-M$ relation as an indicator of the formation
histories of strong lensing BCGs and clusters. 
Our main results are:

\begin{enumerate}

\item  The best-fit cluster mass-temperature relation for our four
dynamically relaxed strong lensing clusters is 
\begin{equation}
\label{eq:mt}
E(z) \left( \frac{M_{2500}} {10^{14} \, h_{70}^{-1} \, M_\odot}
\right) = 1.60 \pm 3.42 \left( \frac{k T}
{5 \, \mathrm{keV}} \right)^{1.43 \pm 1.28} \, ,
\end{equation} 
which is consistent with the theoretical expectation of the $M-T$ relation for relaxed clusters, as well as the $M-T$ relations
determined by other observations and simulations.  We find that
relaxed strong lensing clusters do not deviate from the $M-T$ relation
for the general population of relaxed clusters.

Significantly, we find an inverse correlation between cluster temperature and
cluster concentration that, when incorporated into the
$M-T$ relation, reduces the $M-T$ scatter by a factor of 6, from 360$\%$ to 60$\%$.  By
correcting cluster temperatures according to the
temperature-concentration relation, we find that the
$M-T$ relation given in Equation~\ref{eq:mt} describes not only the
relaxed strong lensing clusters, but the entire cluster population
regardless of dynamical state.  Specifically, the scatter in unrelaxed clusters decreases by a factor of
30, from 500$\%$ in the uncorrected $M-T$ relation to 15$\%$ in the
temperature-corrected $M-T$ relation.  Incorporating concentration
effects into the $M-T$ relation tightens the $M-T$
relation for all clusters, which in turn
reduces the error in the determination of $\sigma_8$ from
cluster counts.  Whereas accurate cluster determinations of $\sigma_8$ were previously made only with relaxed clusters, concentrations enable the inclusion of unrelaxed clusters.  The larger cluster samples possible with the inclusion of unrelaxed clusters offer yet more precise $\sigma_8$ estimates from cluster observations. 

\item The best-fit relation between BCG luminosity and cluster mass for
our sample of strong lensing clusters is
\begin{equation}
\frac{L_\mathrm{K, BCG}} {10^{11} \; h_{70}^{-2} \; L_\odot} = 0.97
\pm 0.17 \left( \frac{M_{200}} {10^{14} \; h_{70}^{-1} \; M_\odot}
\right)^{0.48 \pm 0.09} \; ,
\end{equation}
which is $\sim 85\%$ steeper than the correlations predicted for non-strong-lensing clusters by other
observations, theory, and cosmological simulations.  This result
supports the current evidence that BCGs are built up through mergers
with massive galaxies in other groups and clusters, and also hints that strong
lensing clusters may have more active merging histories than
typical clusters.  A larger sample of strong lensing clusters might enable more definite conclusions about the formation histories of strong lensing clusters.

\end{enumerate}

Accurate cluster mass measurements and full use of the 
range of cluster property interdependencies are key components in the calibration of
clusters as tracers of cosmological parameters.  As we have shown,
gravitational lensing enables the most direct measurements of cluster
mass, without assumptions about the cluster's dynamical state that are
inherent in other methods.  We have also shown that the correlation
between cluster temperature and concentration can significantly reduce the
scatter in the cluster $M-T$ relation, enabling more precise
estimates of $\sigma_8$.  It may be that other cluster scalings can be
effectively combined to reduce the error on additional cosmological
parameter estimates.

\acknowledgements J.M.C. acknowledges support of this work by a National Science
Foundation Graduate Research Fellowship. The work of L.A.M. was
carried out at the Jet Propulsion Laboratory, California
Institute of Technology, with the support of NASA ATFP08-0169. P.N. would like to thank the Radcliffe Institute for Advanced Study and the Center for Astrophysics (CfA)
for providing an intellectually stimulating atmosphere that enabled this work.

\bibliographystyle{apj}
\bibliography{comerford_bcg}

\begin{thebibliography}{85}
\expandafter\ifx\csname natexlab\endcsname\relax\def\natexlab#1{#1}\fi

\bibitem[{Allen(1998)}]{AL98.1}
Allen, S. 1998, MNRAS, 296, 392

\bibitem[{{Allen} {et~al.}(2004){Allen}, {Schmidt}, {Ebeling}, {Fabian}, \&
  {van Speybroeck}}]{AL04.2}
{Allen}, S.~W., {Schmidt}, R.~W., {Ebeling}, H., {Fabian}, A.~C., \& {van
  Speybroeck}, L. 2004, \mnras, 353, 457

\bibitem[{{Allen} {et~al.}(2001){Allen}, {Schmidt}, \& {Fabian}}]{AL01.2}
{Allen}, S.~W., {Schmidt}, R.~W., \& {Fabian}, A.~C. 2001, \mnras, 328, L37

\bibitem[{{Andersson} {et~al.}(2009){Andersson}, {Peterson}, {Madejski}, \&
  {Goobar}}]{AN09.1}
{Andersson}, K., {Peterson}, J.~R., {Madejski}, G., \& {Goobar}, A. 2009, \apj,
  696, 1029

\bibitem[{{Arnaud} {et~al.}(2005){Arnaud}, {Pointecouteau}, \&
  {Pratt}}]{AR05.1}
{Arnaud}, M., {Pointecouteau}, E., \& {Pratt}, G.~W. 2005, \aap, 441, 893

\bibitem[{{Bahcall} \& {Comerford}(2002)}]{BA02.2}
{Bahcall}, N.~A., \& {Comerford}, J.~M. 2002, \apjl, 565, L5

\bibitem[{{Bailey}(1982)}]{BA82.1}
{Bailey}, M.~E. 1982, \mnras, 201, 271

\bibitem[{Bartelmann \& Steinmetz(1996)}]{BA96.2}
Bartelmann, M., \& Steinmetz, M. 1996, MNRAS, 283, 431

\bibitem[{{Bautz} {et~al.}(1982){Bautz}, {Loh}, \& {Wilkinson}}]{BA82.2}
{Bautz}, M., {Loh}, E., \& {Wilkinson}, D.~T. 1982, \apj, 255, 57

\bibitem[{Bertin \& Arnouts(1996)}]{BE96.1}
Bertin, E., \& Arnouts, S. 1996, A\&AS, 117, 393

\bibitem[{{Bildfell} {et~al.}(2008){Bildfell}, {Hoekstra}, {Babul}, \&
  {Mahdavi}}]{BI08.2}
{Bildfell}, C., {Hoekstra}, H., {Babul}, A., \& {Mahdavi}, A. 2008, \mnras,
  389, 1637

\bibitem[{{Borys} {et~al.}(2004){Borys}, {Chapman}, {Donahue}, {Fahlman},
  {Halpern}, {Kneib}, {Newbury}, {Scott}, \& {Smith}}]{BO04.2}
{Borys}, C., {Chapman}, S., {Donahue}, M., {Fahlman}, G., {Halpern}, M.,
  {Kneib}, J.-P., {Newbury}, P., {Scott}, D., \& {Smith}, G.~P. 2004, \mnras,
  352, 759

\bibitem[{{Boylan-Kolchin} {et~al.}(2006){Boylan-Kolchin}, {Ma}, \&
  {Quataert}}]{BO06.1}
{Boylan-Kolchin}, M., {Ma}, C.-P., \& {Quataert}, E. 2006, \mnras, 369, 1081

\bibitem[{Broadhurst {et~al.}(2000)Broadhurst, Huang, Frye, \& Ellis}]{BR00.1}
Broadhurst, T., Huang, X., Frye, B., \& Ellis, R.~S. 2000, ApJ, 534, L15

\bibitem[{{Brough} {et~al.}(2008){Brough}, {Couch}, {Collins}, {Jarrett},
  {Burke}, \& {Mann}}]{BR08.1}
{Brough}, S., {Couch}, W.~J., {Collins}, C.~A., {Jarrett}, T., {Burke}, D.~J.,
  \& {Mann}, R.~G. 2008, \mnras, 385, L103

\bibitem[{{Bruzual} \& {Charlot}(2003)}]{BR03.1}
{Bruzual}, G., \& {Charlot}, S. 2003, \mnras, 344, 1000

\bibitem[{{Bryan} \& {Norman}(1998)}]{BR98.2}
{Bryan}, G.~L., \& {Norman}, M.~L. 1998, \apj, 495, 80

\bibitem[{{Cardiel} {et~al.}(1998){Cardiel}, {Gorgas}, \&
  {Aragon-Salamanca}}]{CA98.1}
{Cardiel}, N., {Gorgas}, J., \& {Aragon-Salamanca}, A. 1998, \mnras, 298, 977

\bibitem[{{Churazov} {et~al.}(2001){Churazov}, {Br{\"u}ggen}, {Kaiser},
  {B{\"o}hringer}, \& {Forman}}]{CH01.1}
{Churazov}, E., {Br{\"u}ggen}, M., {Kaiser}, C.~R., {B{\"o}hringer}, H., \&
  {Forman}, W. 2001, \apj, 554, 261

\bibitem[{{Cole} {et~al.}(2000){Cole}, {Lacey}, {Baugh}, \& {Frenk}}]{CO00.1}
{Cole}, S., {Lacey}, C.~G., {Baugh}, C.~M., \& {Frenk}, C.~S. 2000, \mnras,
  319, 168

\bibitem[{{Comerford} {et~al.}(2006){Comerford}, {Meneghetti}, {Bartelmann}, \&
  {Schirmer}}]{CO06.1}
{Comerford}, J.~M., {Meneghetti}, M., {Bartelmann}, M., \& {Schirmer}, M. 2006,
  \apj, 642, 39

\bibitem[{{Comerford} \& {Natarajan}(2007)}]{CO07.1}
{Comerford}, J.~M., \& {Natarajan}, P. 2007, \mnras, 379, 190

\bibitem[{{Cooray} \& {Milosavljevi{\'c}}(2005)}]{CO05.1}
{Cooray}, A., \& {Milosavljevi{\'c}}, M. 2005, \apjl, 627, L85

\bibitem[{{Cotter} {et~al.}(2002){Cotter}, {Buttery}, {Das}, {Jones},
  {Grainge}, {Pooley}, \& {Saunders}}]{CO02.1}
{Cotter}, G., {Buttery}, H.~J., {Das}, R., {Jones}, M.~E., {Grainge}, K.,
  {Pooley}, G.~G., \& {Saunders}, R. 2002, \mnras, 334, 323

\bibitem[{{Cowie} \& {Binney}(1977)}]{CO77.1}
{Cowie}, L.~L., \& {Binney}, J. 1977, \apj, 215, 723

\bibitem[{{Crawford} {et~al.}(1999){Crawford}, {Allen}, {Ebeling}, {Edge}, \&
  {Fabian}}]{CR99.1}
{Crawford}, C.~S., {Allen}, S.~W., {Ebeling}, H., {Edge}, A.~C., \& {Fabian},
  A.~C. 1999, \mnras, 306, 857

\bibitem[{{Czoske} {et~al.}(2002){Czoske}, {Moore}, {Kneib}, \&
  {Soucail}}]{CZ02.1}
{Czoske}, O., {Moore}, B., {Kneib}, J.-P., \& {Soucail}, G. 2002, \aap, 386, 31

\bibitem[{{De Filippis} {et~al.}(2003){De Filippis}, {Schindler}, \&
  {Castillo-Morales}}]{DE03.1}
{De Filippis}, E., {Schindler}, S., \& {Castillo-Morales}, A. 2003, \aap, 404,
  63

\bibitem[{{Donahue} {et~al.}(1999){Donahue}, {Voit}, {Scharf}, {Gioia},
  {Mullis}, {Hughes}, \& {Stocke}}]{DO99.1}
{Donahue}, M., {Voit}, G.~M., {Scharf}, C.~A., {Gioia}, I.~M., {Mullis}, C.~R.,
  {Hughes}, J.~P., \& {Stocke}, J.~T. 1999, \apj, 527, 525

\bibitem[{{Dubinski}(1998)}]{DU98.1}
{Dubinski}, J. 1998, \apj, 502, 141

\bibitem[{{Ellingson} {et~al.}(1998){Ellingson}, {Yee}, {Abraham}, {Morris}, \&
  {Carlberg}}]{EL98.1}
{Ellingson}, E., {Yee}, H.~K.~C., {Abraham}, R.~G., {Morris}, S.~L., \&
  {Carlberg}, R.~G. 1998, \apjs, 116, 247

\bibitem[{{Elston} {et~al.}(2006){Elston}, {Gonzalez}, {McKenzie}, {Brodwin},
  {Brown}, {Cardona}, {Dey}, {Dickinson}, {Eisenhardt}, {Jannuzi}, {Lin},
  {Mohr}, {Raines}, {Stanford}, \& {Stern}}]{EL06.1}
{Elston}, R.~J., {Gonzalez}, A.~H., {McKenzie}, E., {Brodwin}, M., {Brown},
  M.~J.~I., {Cardona}, G., {Dey}, A., {Dickinson}, M., {Eisenhardt}, P.~R.,
  {Jannuzi}, B.~T., {Lin}, Y.-T., {Mohr}, J.~J., {Raines}, S.~N., {Stanford},
  S.~A., \& {Stern}, D. 2006, \apj, 639, 816

\bibitem[{{Evrard} {et~al.}(2002){Evrard}, {MacFarland}, {Couchman}, {Colberg},
  {Yoshida}, {White}, {Jenkins}, {Frenk}, {Pearce}, {Peacock}, \&
  {Thomas}}]{EV02.1}
{Evrard}, A.~E., {MacFarland}, T.~J., {Couchman}, H.~M.~P., {Colberg}, J.~M.,
  {Yoshida}, N., {White}, S.~D.~M., {Jenkins}, A., {Frenk}, C.~S., {Pearce},
  F.~R., {Peacock}, J.~A., \& {Thomas}, P.~A. 2002, \apj, 573, 7

\bibitem[{{Evrard} {et~al.}(1996){Evrard}, {Metzler}, \& {Navarro}}]{EV96.1}
{Evrard}, A.~E., {Metzler}, C.~A., \& {Navarro}, J.~F. 1996, \apj, 469, 494

\bibitem[{{Finoguenov} {et~al.}(2001){Finoguenov}, {Reiprich}, \&
  {B{\"o}hringer}}]{FI01.1}
{Finoguenov}, A., {Reiprich}, T.~H., \& {B{\"o}hringer}, H. 2001, \aap, 368,
  749

\bibitem[{{Gavazzi}(2005)}]{GA05.1}
{Gavazzi}, R. 2005, in IAU Symposium, 179--184

\bibitem[{{Haiman} {et~al.}(2001){Haiman}, {Mohr}, \& {Holder}}]{HA01.2}
{Haiman}, Z., {Mohr}, J.~J., \& {Holder}, G.~P. 2001, \apj, 553, 545

\bibitem[{{Hausman} \& {Ostriker}(1978)}]{HA78.1}
{Hausman}, M.~A., \& {Ostriker}, J.~P. 1978, \apj, 224, 320

\bibitem[{{Henry}(2004)}]{HE04.2}
{Henry}, J.~P. 2004, \apj, 609, 603

\bibitem[{{Hicks} \& {Mushotzky}(2005)}]{HI05.1}
{Hicks}, A.~K., \& {Mushotzky}, R. 2005, \apjl, 635, L9

\bibitem[{{Hoekstra}(2007)}]{HO07.2}
{Hoekstra}, H. 2007, \mnras, 379, 317

\bibitem[{{Horner}(2001)}]{HO01.1}
{Horner}, D.~J. 2001, PhD thesis, University of Maryland College Park

\bibitem[{{Jeltema} {et~al.}(2008){Jeltema}, {Hallman}, {Burns}, \&
  {Motl}}]{JE08.1}
{Jeltema}, T.~E., {Hallman}, E.~J., {Burns}, J.~O., \& {Motl}, P.~M. 2008,
  \apj, 681, 167

\bibitem[{{Kay} {et~al.}(2005){Kay}, {da Silva}, {Aghanim}, {Blanchard},
  {Liddle}, {Puget}, {Sadat}, \& {Thomas}}]{KA05.1}
{Kay}, S.~T., {da Silva}, A.~C., {Aghanim}, N., {Blanchard}, A., {Liddle},
  A.~R., {Puget}, J.-L., {Sadat}, R., \& {Thomas}, P.~A. 2005, Advances in
  Space Research, 36, 694

\bibitem[{{Kneib} {et~al.}(2003){Kneib}, {Hudelot}, {Ellis}, {Treu}, {Smith},
  {Marshall}, {Czoske}, {Smail}, \& {Natarajan}}]{KN03.1}
{Kneib}, J., {Hudelot}, P., {Ellis}, R.~S., {Treu}, T., {Smith}, G.~P.,
  {Marshall}, P., {Czoske}, O., {Smail}, I., \& {Natarajan}, P. 2003, ApJ, 598,
  804

\bibitem[{Kneib {et~al.}(1993)Kneib, Mellier, Fort, \& Mathez}]{KN93.1}
Kneib, J., Mellier, Y., Fort, B., \& Mathez, G. 1993, A\&A, 273, 367

\bibitem[{{Kravtsov} {et~al.}(2006){Kravtsov}, {Vikhlinin}, \&
  {Nagai}}]{KR06.1}
{Kravtsov}, A.~V., {Vikhlinin}, A., \& {Nagai}, D. 2006, \apj, 650, 128

\bibitem[{{Levine} {et~al.}(2002){Levine}, {Schulz}, \& {White}}]{LE02.1}
{Levine}, E.~S., {Schulz}, A.~E., \& {White}, M. 2002, \apj, 577, 569

\bibitem[{{Lin} \& {Mohr}(2004)}]{LI04.3}
{Lin}, Y.-T., \& {Mohr}, J.~J. 2004, \apj, 617, 879

\bibitem[{{Mahdavi} {et~al.}(2008){Mahdavi}, {Hoekstra}, {Babul}, \&
  {Henry}}]{MA08.1}
{Mahdavi}, A., {Hoekstra}, H., {Babul}, A., \& {Henry}, J.~P. 2008, \mnras,
  384, 1567

\bibitem[{{Mathiesen} \& {Evrard}(2001)}]{MA01.2}
{Mathiesen}, B.~F., \& {Evrard}, A.~E. 2001, \apj, 546, 100

\bibitem[{{Maughan} {et~al.}(2008){Maughan}, {Jones}, {Forman}, \& {Van
  Speybroeck}}]{MA08.2}
{Maughan}, B.~J., {Jones}, C., {Forman}, W., \& {Van Speybroeck}, L. 2008,
  \apjs, 174, 117

\bibitem[{{Mazzotta} {et~al.}(2004){Mazzotta}, {Rasia}, {Moscardini}, \&
  {Tormen}}]{MA04.1}
{Mazzotta}, P., {Rasia}, E., {Moscardini}, L., \& {Tormen}, G. 2004, \mnras,
  354, 10

\bibitem[{{McNamara} {et~al.}(2006){McNamara}, {Rafferty}, {B{\^i}rzan},
  {Steiner}, {Wise}, {Nulsen}, {Carilli}, {Ryan}, \& {Sharma}}]{MC06.1}
{McNamara}, B.~R., {Rafferty}, D.~A., {B{\^i}rzan}, L., {Steiner}, J., {Wise},
  M.~W., {Nulsen}, P.~E.~J., {Carilli}, C.~L., {Ryan}, R., \& {Sharma}, M.
  2006, \apj, 648, 164

\bibitem[{{Mellier} {et~al.}(1994){Mellier}, {Fort}, {Bonnet}, \&
  J.P.}]{ME94.1}
{Mellier}, Y., {Fort}, B., {Bonnet}, H., \& J.P., K. 1994, in "Cosmological
  Aspects of X-ray Clusters of Galaxies", W.C. Seitter ed., NATO ASI Series,
  441, 219

\bibitem[{{Merritt}(1985)}]{ME85.1}
{Merritt}, D. 1985, \apj, 289, 18

\bibitem[{{Mohr} {et~al.}(1993){Mohr}, {Fabricant}, \& {Geller}}]{MO93.1}
{Mohr}, J.~J., {Fabricant}, D.~G., \& {Geller}, M.~J. 1993, \apj, 413, 492

\bibitem[{{Moran} {et~al.}(2005){Moran}, {Ellis}, {Treu}, {Smail}, {Dressler},
  {Coil}, \& {Smith}}]{MO05.1}
{Moran}, S.~M., {Ellis}, R.~S., {Treu}, T., {Smail}, I., {Dressler}, A.,
  {Coil}, A.~L., \& {Smith}, G.~P. 2005, \apj, 634, 977

\bibitem[{{Natarajan} {et~al.}(2009){Natarajan}, {Kneib}, {Smail}, {Treu},
  {Ellis}, {Moran}, {Limousin}, \& {Czoske}}]{NA09.1}
{Natarajan}, P., {Kneib}, J.-P., {Smail}, I., {Treu}, T., {Ellis}, R., {Moran},
  S., {Limousin}, M., \& {Czoske}, O. 2009, \apj, 693, 970

\bibitem[{Navarro {et~al.}(1996)Navarro, Frenk, \& White}]{NA96.1}
Navarro, J., Frenk, C., \& White, S. 1996, ApJ, 462, 563

\bibitem[{Navarro {et~al.}(1997)Navarro, Frenk, \& White}]{NA97.1}
---. 1997, ApJ, 490, 493

\bibitem[{{Nevalainen} {et~al.}(2000){Nevalainen}, {Markevitch}, \&
  {Forman}}]{NE00.2}
{Nevalainen}, J., {Markevitch}, M., \& {Forman}, W. 2000, \apj, 532, 694

\bibitem[{{Ostriker} \& {Tremaine}(1975)}]{OS75.1}
{Ostriker}, J.~P., \& {Tremaine}, S.~D. 1975, \apjl, 202, L113

\bibitem[{{Ota} \& {Mitsuda}(2004)}]{OT04.2}
{Ota}, N., \& {Mitsuda}, K. 2004, \aap, 428, 757

\bibitem[{{Ota} {et~al.}(1998){Ota}, {Mitsuda}, \& {Fukazawa}}]{OT98.1}
{Ota}, N., {Mitsuda}, K., \& {Fukazawa}, Y. 1998, ApJ, 495, 170

\bibitem[{{Ota} {et~al.}(2000){Ota}, {Mitsuda}, {Hattori}, \&
  {Mihara}}]{OT00.1}
{Ota}, N., {Mitsuda}, K., {Hattori}, M., \& {Mihara}, T. 2000, \apj, 530, 172

\bibitem[{{Ota} {et~al.}(2004){Ota}, {Pointecouteau}, {Hattori}, \&
  {Mitsuda}}]{OT04.1}
{Ota}, N., {Pointecouteau}, E., {Hattori}, M., \& {Mitsuda}, K. 2004, \apj,
  601, 120

\bibitem[{{Pierpaoli} {et~al.}(2003){Pierpaoli}, {Borgani}, {Scott}, \&
  {White}}]{PI03.1}
{Pierpaoli}, E., {Borgani}, S., {Scott}, D., \& {White}, M. 2003, \mnras, 342,
  163

\bibitem[{{Schindler} {et~al.}(1998){Schindler}, {Belloni}, {Ikebe}, {Hattori},
  {Wambsganss}, \& {Tanaka}}]{SC98.3}
{Schindler}, S., {Belloni}, P., {Ikebe}, Y., {Hattori}, M., {Wambsganss}, J.,
  \& {Tanaka}, Y. 1998, \aap, 338, 843

\bibitem[{{Schindler} \& {Wambsganss}(1996)}]{SC96.3}
{Schindler}, S., \& {Wambsganss}, J. 1996, \aap, 313, 113

\bibitem[{{Schmidt} \& {Allen}(2007)}]{SC07.2}
{Schmidt}, R.~W., \& {Allen}, S.~W. 2007, \mnras, 379, 209

\bibitem[{{Schuecker} {et~al.}(2003){Schuecker}, {B{\"o}hringer}, {Collins}, \&
  {Guzzo}}]{SC03.2}
{Schuecker}, P., {B{\"o}hringer}, H., {Collins}, C.~A., \& {Guzzo}, L. 2003,
  \aap, 398, 867

\bibitem[{{Smith} {et~al.}(2005){Smith}, {Kneib}, {Smail}, {Mazzotta},
  {Ebeling}, \& {Czoske}}]{SM05.1}
{Smith}, G.~P., {Kneib}, J., {Smail}, I., {Mazzotta}, P., {Ebeling}, H., \&
  {Czoske}, O. 2005, \mnras, 359, 417

\bibitem[{{Somerville} \& {Primack}(1999)}]{SO99.1}
{Somerville}, R.~S., \& {Primack}, J.~R. 1999, \mnras, 310, 1087

\bibitem[{{Vale} \& {Ostriker}(2006)}]{VA06.1}
{Vale}, A., \& {Ostriker}, J.~P. 2006, \mnras, 371, 1173

\bibitem[{{Vikhlinin} {et~al.}(2006){Vikhlinin}, {Kravtsov}, {Forman}, {Jones},
  {Markevitch}, {Murray}, \& {Van Speybroeck}}]{VI06.1}
{Vikhlinin}, A., {Kravtsov}, A., {Forman}, W., {Jones}, C., {Markevitch}, M.,
  {Murray}, S.~S., \& {Van Speybroeck}, L. 2006, \apj, 640, 691

\bibitem[{{Vikhlinin} {et~al.}(2009){Vikhlinin}, {Kravtsov}, {Burenin},
  {Ebeling}, {Forman}, {Hornstrup}, {Jones}, {Murray}, {Nagai}, {Quintana}, \&
  {Voevodkin}}]{VI09.1}
{Vikhlinin}, A., {Kravtsov}, A.~V., {Burenin}, R.~A., {Ebeling}, H., {Forman},
  W.~R., {Hornstrup}, A., {Jones}, C., {Murray}, S.~S., {Nagai}, D.,
  {Quintana}, H., \& {Voevodkin}, A. 2009, \apj, 692, 1060

\bibitem[{{Voit}(2005)}]{VO05.1}
{Voit}, G.~M. 2005, Reviews of Modern Physics, 77, 207

\bibitem[{Wechsler {et~al.}(2002)Wechsler, Bullock, Primack, Kravtsov, \&
  Dekel}]{WE02.1}
Wechsler, R., Bullock, J., Primack, J., Kravtsov, A., \& Dekel, A. 2002, ApJ,
  568, 52

\bibitem[{{Worrall} {et~al.}(2001){Worrall}, {Birkinshaw}, {Hardcastle}, \&
  {Lawrence}}]{WO01.1}
{Worrall}, D.~M., {Birkinshaw}, M., {Hardcastle}, M.~J., \& {Lawrence}, C.~R.
  2001, \mnras, 326, 1127

\bibitem[{{Wu}(2000)}]{WU00.1}
{Wu}, X.-P. 2000, \mnras, 316, 299

\bibitem[{{Yamada} {et~al.}(2002){Yamada}, {Koyama}, {Nakata}, {Kajisawa},
  {Tanaka}, {Kodama}, {Okamura}, \& {De Propris}}]{YA02.1}
{Yamada}, T., {Koyama}, Y., {Nakata}, F., {Kajisawa}, M., {Tanaka}, I.,
  {Kodama}, T., {Okamura}, S., \& {De Propris}, R. 2002, \apjl, 577, L89

\bibitem[{{Zhang} {et~al.}(2005){Zhang}, {B{\"o}hringer}, {Mellier}, {Soucail},
  \& {Forman}}]{ZH05.2}
{Zhang}, Y., {B{\"o}hringer}, H., {Mellier}, Y., {Soucail}, G., \& {Forman}, W.
  2005, \aap, 429, 85

\bibitem[{{Zhang} {et~al.}(2008){Zhang}, {Finoguenov}, {B{\"o}hringer},
  {Kneib}, {Smith}, {Kneissl}, {Okabe}, \& {Dahle}}]{ZH08.1}
{Zhang}, Y., {Finoguenov}, A., {B{\"o}hringer}, H., {Kneib}, J., {Smith},
  G.~P., {Kneissl}, R., {Okabe}, N., \& {Dahle}, H. 2008, \aap, 482, 451

\bibitem[{{Zhang} {et~al.}(2009){Zhang}, {Reiprich}, {Finoguenov}, {Hudson}, \&
  {Sarazin}}]{ZH09.1}
{Zhang}, Y., {Reiprich}, T.~H., {Finoguenov}, A., {Hudson}, D.~S., \&
  {Sarazin}, C.~L. 2009, \apj, 699, 1178

\end{thebibliography}

\end{document}